# ON THE CHARACTERISTIC FORM OF HISTOGRAMS APPEARING AT THE CULMINATION OF SOLAR ECLIPSE

S.E. Shnoll[1,2], V.A. Panchelyuga[2]

Lomonosov's Moscow State University, Moscow, Russia (1)
Institute of Theoretical and Experimental Biophysics RAS, Pushchino, Russia (2)
*shnoll@iteb.ru, panvic333@yahoo.com*

As shown in a number of our works, the form of histograms – distributions of amplitude fluctuations – varies regularly in time, with these variations being similar for processes of any nature, from biochemical reactions to noise in the gravitational antenna and all types of the radioactive decay [1-5]. In particular, we have revealed basic laws, suggesting a cosmo-physical nature of these phenomena, in the time series created by the noise generators of the global GCP net [7]. On the basis of all the results obtained, a conclusion has been made that the histogram form is determined by fluctuations of the space-time, which depend on the movement of the measured system ("laboratory") relative to the heavenly bodies [8].

An important step to understand the nature of these phenomena was the finding that at the moments of the new Moon, a specific histogram form appears practically simultaneously at different geographical points, from Arctic to Antarctic, in middle latitudes of West and East hemispheres [9]. This effect seems to be not due to a change of the tide-generating forces; to explain it, nontrivial hypotheses are needed.

The present paper shows the appearance of specific histogram forms at the culminations of the solar eclipses (moments of the geocentric superposition of the Sun and the Moon), with the "eclipse" forms differing from the "new-moon" ones. Specific histogram forms appear practically simultaneously "all over the Earth" and depend nor on the geographical coordinates, nor on the nature of the process studied.

## 1. Introduction
As shown in our works, the fine structure of sampling distributions (the form of the corresponding histograms constructed from the results of the successive measurements of processes of various nature) is determined by cosmo-physical factors, i.e. by the position of the measured object relative to the Sun, the Moon and the sphere of immobile stars [1-8]. On the new moon, histograms with a characteristic form are observed simultaneously at different geographical points [9]. This paper considers analogous results, indicating a characteristic histogram form to appear during the solar eclipses.

In the time series studied, we have not revealed regular changes of the radioactive decay rate during solar eclipses. However, we have found that changing regularly at these moments are the forms of histograms constructed from those time series. This suggests, among other things, that the analysis of histogram form is a more sensitive indicator for various effects than the analysis of time series themselves.

## 2. Materials and Methods
The main method used in this work was a comparison of the histogram form constructed from a small number (30-60) of the results of successive measurements. We measured alpha-decay of $^{239}$Pu samples, which were immobilized on semiconductor detectors in devices designed by I.A. Rubinstein [10]. Such measurements, with the number of decays counted every second, have been carried out in our laboratory uninterruptedly, whenever possible, for many years. So now we have data for nine solar eclipses, which happened within 2000-2005.

Another source of data used in this work is time series created by the noise generators of the global GCP network. These investigations are given in a special work [7].

The techniques of measurements, histogram construction and analysis of the results obtained are described in papers published earlier [1-6, 10].

## 3. Results

The investigation of processes of various nature through comparison of histograms allows one to discover regularities that are difficult to be revealed by traditional statistic methods. The study of effects of the solar eclipse was no exception: here the advantages of histogram analysis were also apparent. To demonstrate this, we shall consider in details the results of measurements of $^{239}$Pu alpha-activity, which were carried out in Pushchino during the solar eclipse of April 8-9, 2005.

Table 1 and Fig. 1 show a segment of a time series, containing the results of one-second measurements obtained during the solar eclipse of April 8-9, 2005. The interval, corresponding to the eclipse culmination, is marked with bold red.

Table 1.

| no. | cps | no. | cps | no. | cps | no. | cps | no. | cps | no. | cps | no. | cps | no. | cps | no. | cps |
|---|---|---|---|---|---|---|---|---|---|---|---|---|---|---|---|---|---|
| 1 | 304 | 21 | 252 | 41 | **283** | 61 | **286** | 81 | **283** | 101 | 299 | 121 | 260 | 141 | 271 | 161 | 265 |
| 2 | 281 | 22 | 296 | 42 | **259** | 62 | **303** | 82 | **280** | 102 | 271 | 122 | 268 | 142 | 291 | 162 | 267 |
| 3 | 271 | 23 | 291 | 43 | **275** | 63 | **245** | 83 | **305** | 103 | 295 | 123 | 275 | 143 | 302 | 163 | 278 |
| 4 | 257 | 24 | 265 | 44 | **276** | 64 | **276** | 84 | **287** | 104 | 267 | 124 | 267 | 144 | 281 | 164 | 260 |
| 5 | 265 | 25 | 241 | 45 | **309** | 65 | **304** | 85 | **290** | 105 | 285 | 125 | 307 | 145 | 270 | 165 | 271 |
| 6 | 288 | 26 | 291 | 46 | **297** | 66 | **275** | 86 | **288** | 106 | 270 | 126 | 254 | 146 | 242 | 166 | 280 |
| 7 | 276 | 27 | 259 | 47 | **279** | 67 | **285** | 87 | **255** | 107 | 278 | 127 | 260 | 147 | 286 | 167 | 264 |
| 8 | 293 | 28 | 288 | 48 | **306** | 68 | **287** | 88 | **269** | 108 | 298 | 128 | 278 | 148 | 287 | 168 | 258 |
| 9 | 274 | 29 | 269 | 49 | **262** | 69 | **299** | 89 | **298** | 109 | 311 | 129 | 297 | 149 | 273 | 169 | 263 |
| 10 | 286 | 30 | 275 | 50 | **270** | 70 | **248** | 90 | **294** | 110 | 268 | 130 | 289 | 150 | 252 | 170 | 270 |
| 11 | 279 | 31 | **313** | 51 | **265** | 71 | **281** | 91 | 269 | 111 | 290 | 131 | 275 | 151 | 297 | 171 | 264 |
| 12 | 270 | 32 | **336** | 52 | **286** | 72 | **276** | 92 | 284 | 112 | 290 | 132 | 294 | 152 | 258 | 172 | 316 |
| 13 | 274 | 33 | **250** | 53 | **257** | 73 | **292** | 93 | 257 | 113 | 294 | 133 | 266 | 153 | 282 | 173 | 276 |
| 14 | 288 | 34 | **264** | 54 | **291** | 74 | **284** | 94 | 300 | 114 | 297 | 134 | 280 | 154 | 261 | 174 | 298 |
| 15 | 269 | 35 | **311** | 55 | **269** | 75 | **287** | 95 | 313 | 115 | 278 | 135 | 284 | 155 | 277 | 175 | 287 |
| 16 | 287 | 36 | **288** | 56 | **270** | 76 | **265** | 96 | 270 | 116 | 290 | 136 | 244 | 156 | 254 | 176 | 274 |
| 17 | 296 | 37 | **295** | 57 | **292** | 77 | **266** | 97 | 280 | 117 | 266 | 137 | 289 | 157 | 256 | 177 | 279 |
| 18 | 298 | 38 | **282** | 58 | **271** | 78 | **279** | 98 | 300 | 118 | 259 | 138 | 264 | 158 | 268 | 178 | 280 |
| 19 | 264 | 39 | **271** | 59 | **304** | 79 | **273** | 99 | 312 | 119 | 324 | 139 | 316 | 159 | 276 | 179 | 271 |
| 20 | 289 | 40 | **301** | 60 | **263** | 80 | **269** | 100 | 299 | 120 | 304 | 140 | 266 | 160 | 288 | 180 | 276 |

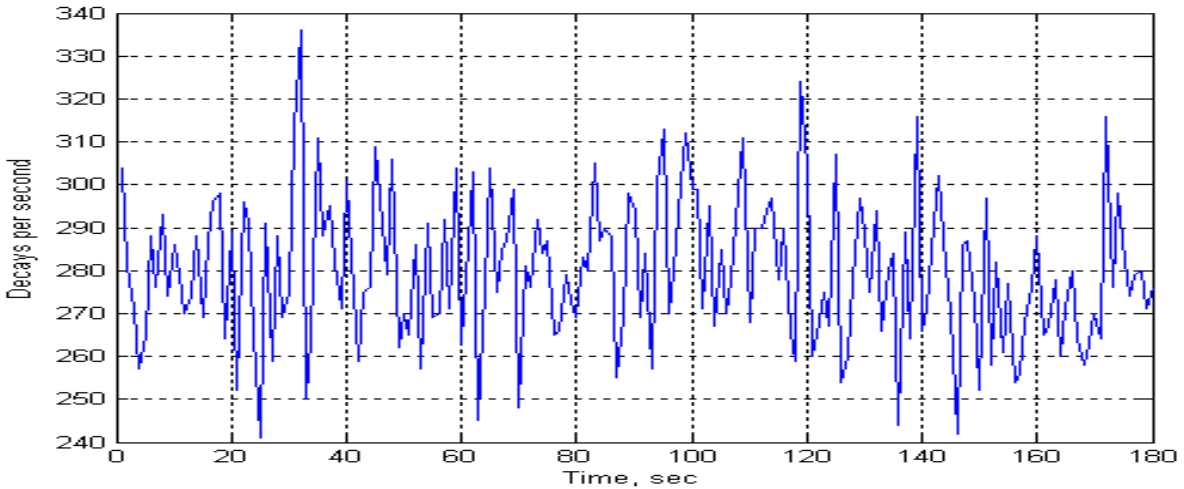

Fig. 1. A segment of a time series, representing the results of measurement of $^{239}$Pu alpha-activity during the solar eclipse of April 9, 2005 at 00:43 by Moscow summer time. Along the X-axis is the time of measurement, sec; along the Y-axis is the number of $^{239}$Pu decays per second. The eclipse culmination falls in the interval of 31-90 sec (see Table 1).

Obviously, it is difficult to reveal any regular changes in the rate of radioactive decay during the culmination of the solar eclipse. At the same time, the histogram corresponding to the eclipse culmination has a particular form.

Some histograms constructed from these measurements are represented in Fig. 2. Each histogram is constructed from the results of 60 one-second measurements, i.e. with the total time of 1 min. The histograms are smoothed for 11 times by a four-point rectangular window. The order number of a histogram corresponds to the number of minutes passed from the beginning of measurement. Histograms No. 42, 43 and 44 correspond to the segment in Fig. 1. Corresponding to the eclipse culmination is histogram No. 43 emphasized with a bold frame. We will demonstrate further that histogram No. 43 has a form specific for the moments of eclipse culminations.

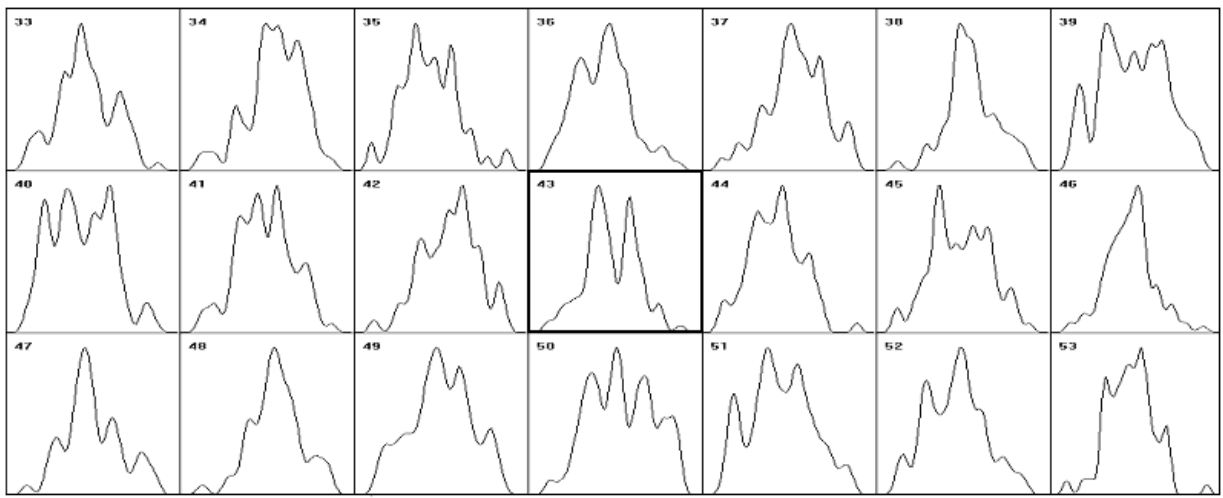

Fig. 2. A fragment of the computer journal representing a sequence of histograms, each of them constructed from 60 one-second measurements of 239Pu alpha-activity. The histogram order numbers correspond to the number of minutes passing from 00:00 (by Moscow winter time) of April 9, 2005. A bold frame borders one-minute histogram No. 43 constructed from the results of measurements during the eclipse culmination. Along the X-axis of each histogram is the rate of 239Pu decay; along the Y-axis is the number of measurements corresponding to certain rates of decay.

When histograms are constructed from 30-point segments, i.e. from measurements lasted 0.5 min, the single "anomalous" histogram No. 43 is replaced with two histograms: No. 85 and 86, which also have specific form (Fig. 3). These two histograms have mirror symmetry: they are superimposed after rotation one of them around vertical axis (Fig. 4).

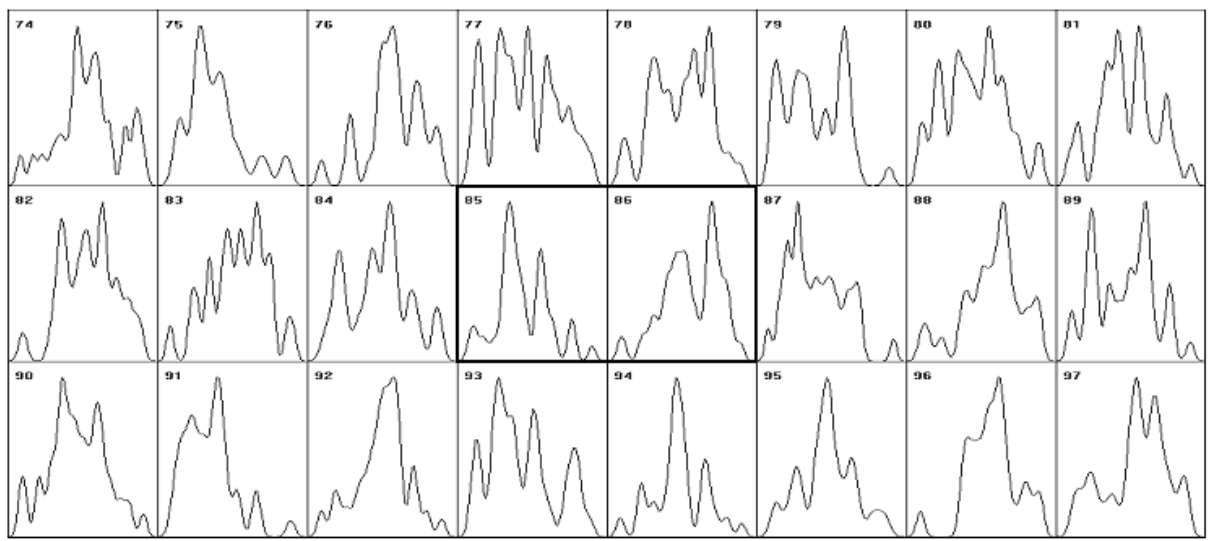

Fig. 3. Characteristic mirror-symmetric forms of two 0.5-min histograms: No. 85 and No. 86, which are constructed from the results of measurements during the culmination of the eclipse of April 9, 2005.

The main thesis of our work is the following. The form specific for histogram No. 43, corresponding to the culmination of the solar eclipse of April 9, 2005 (Fig. 2), is also observed at the study of other solar eclipses, by using of fluctuations measurements in different processes of various nature. This form appears synchronously at different geographical points, independently of the eclipse visibility at the place where the measurements are carried out.

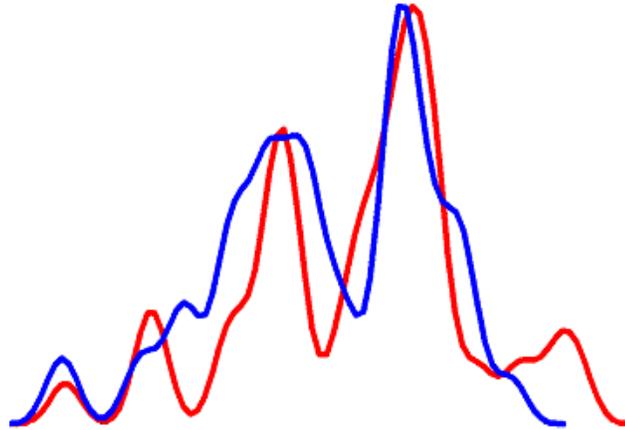

Fig. 4. The forms of the two 0.5-min histograms (No. 85 and 86) constructed from the results of measurements during the culmination of the eclipse of April 9, 2005 are mirror-symmetric, i.e. similar when superimposed after rotation of one of them around its vertical axis.

This thesis is illustrated with a number of charts (Fig. 5-16) which show, in the same way as Fig. 2, sets of histograms corresponding to the culminations of solar eclipses upon examination of various processes, in different years and at different geographical locations. In these figures, histograms are constructed from the results of measurements of alpha-activity of $^{239}$Pu samples (Fig. 5-7 and Fig. 14-16) or measurements of noise in the generators of the GCP system (Fig. 8-13). In each figure, there is a histogram in the center of the second row, corresponding to the culminations of the solar eclipse. All figures show that at these culmination points, histograms of the same specific form are observed at different geographical locations simultaneously (with an accuracy of 0.5-1 min) and irrespectively of the nature of the process measured.

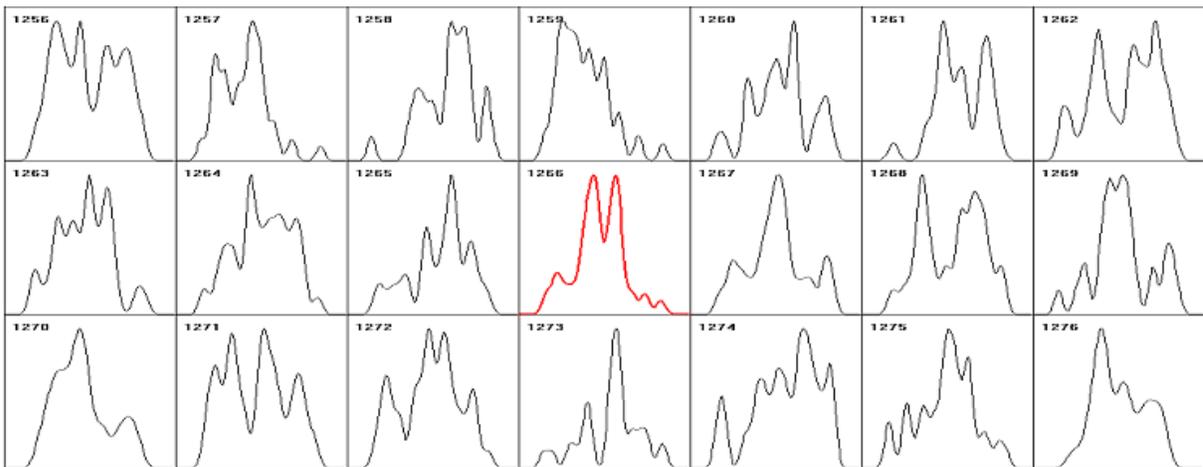

Fig. 5. 0.5-min histograms constructed from the results of measurements of $^{239}$Pu alpha-activity during the solar eclipse of December 4, 2002 in Athens (in latitude 38° north and longitude 23°66′ east). Histogram No. 1266 (at the center of the second row) corresponds to the eclipse culmination with an accuracy of 0.5 min.

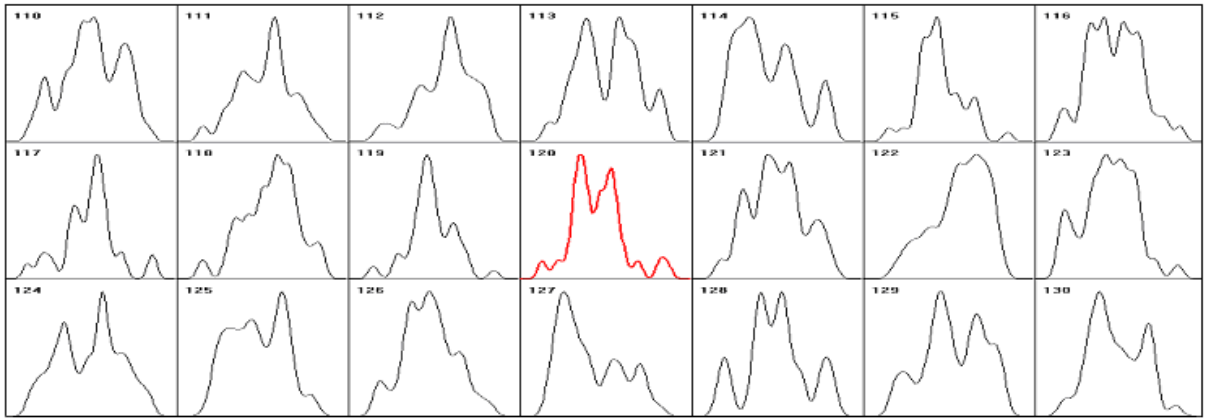

Fig. 6. 1-min histograms constructed from the results of measurements of $^{239}$Pu alpha-activity in the experiment with a collimator directed at the Pole Star during the solar eclipse of December 4, 2002 in Pushchino (in latitude 54°8′ north and longitude 37°6′ east). Histogram No. 120 (at the center of the second row) corresponds to the eclipse culmination with an accuracy of 1 min.

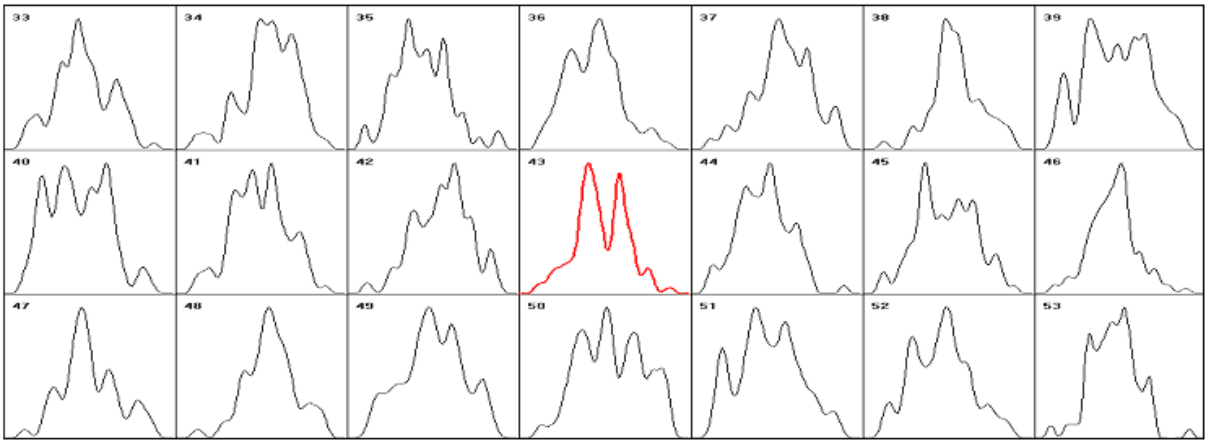

Fig. 7. 1-min histograms constructed from the results of measurements of $^{239}$Pu alpha-activity in the experiments with a collimator directed at the Sun and rotating clockwise (1 rotation per day) during the solar eclipse of April 9, 2005 in Pushchino (in latitude 54°8′ north and longitude 37°6′ east). Histogram No. 43 (at the center of the second row) corresponds to the eclipse culmination with an accuracy of 1 min.

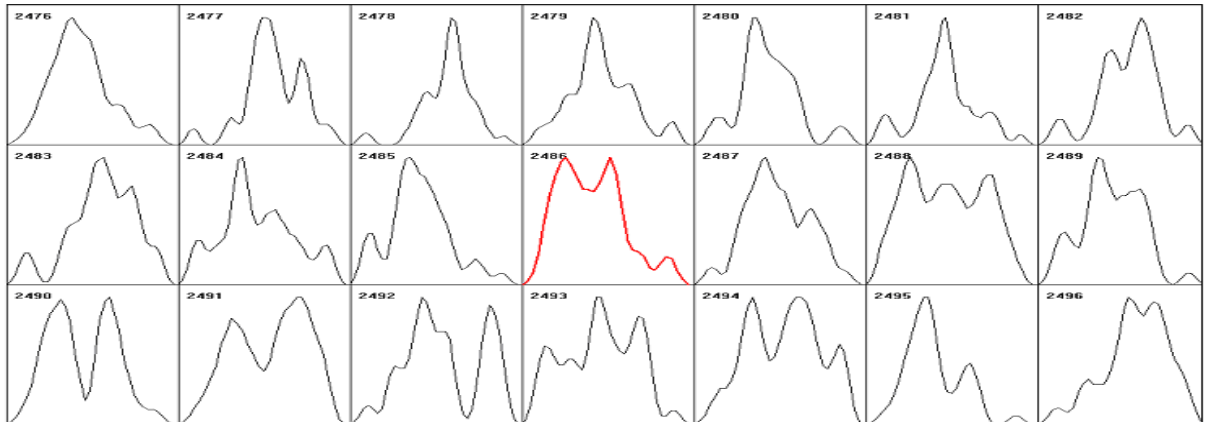

Fig. 8. 0.5-min histograms constructed from the results of measurements of noise in GCP-generator No. 28 (Princeton NJ, USA; in latitude 40°35′ north and longitude 74°66′ west) during the solar eclipse of April 8, 2005. Histogram No. 2486 (at the center of the second row) corresponds to the eclipse culmination with an accuracy of 0.5 min.

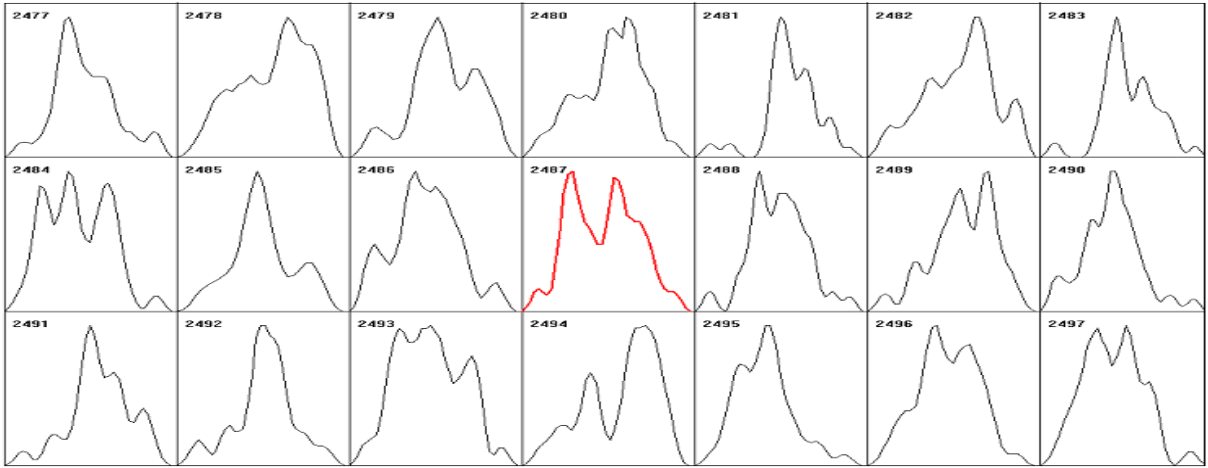

Fig. 9. 0.5-min histograms constructed from the results of measurements of noise in GCP-generator No. 37 (Neuchbtel, Switzerland; in latitude 47°08′ north and longitude 7°06′ west) during the solar eclipse of April 8, 2005. Histogram No. 2487 (at the center of the second row) appears 0.5 min later the eclipse culmination.

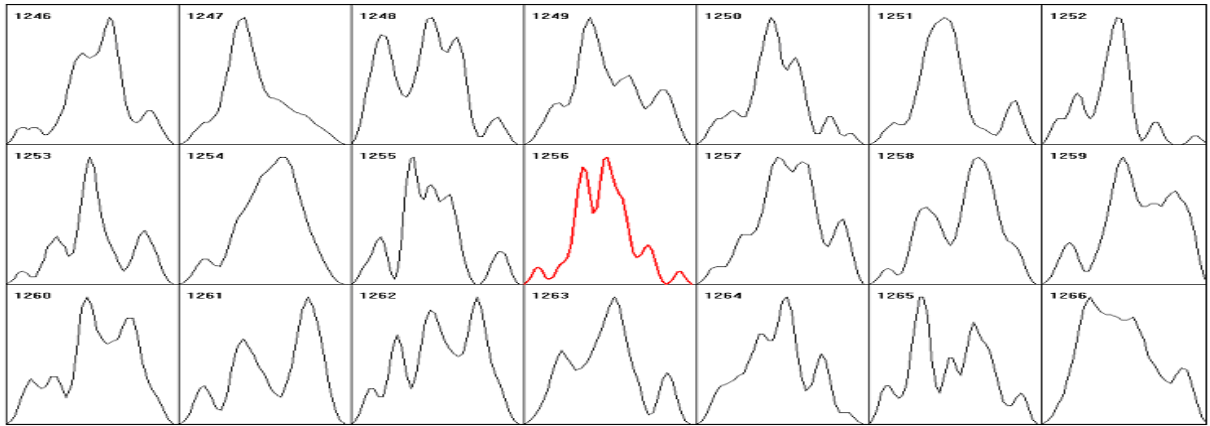

Fig. 10. 0.5-min histograms constructed from the results of measurements of noise in GCP-generator No. 100 (Suva, Fiji; in latitude 17°75′ south and longitude 177°45′ west) during the solar eclipse of October 3, 2005. Histogram No. 1256 (at the center of the second row) corresponds to the eclipse culmination with an accuracy of 0.5 min.

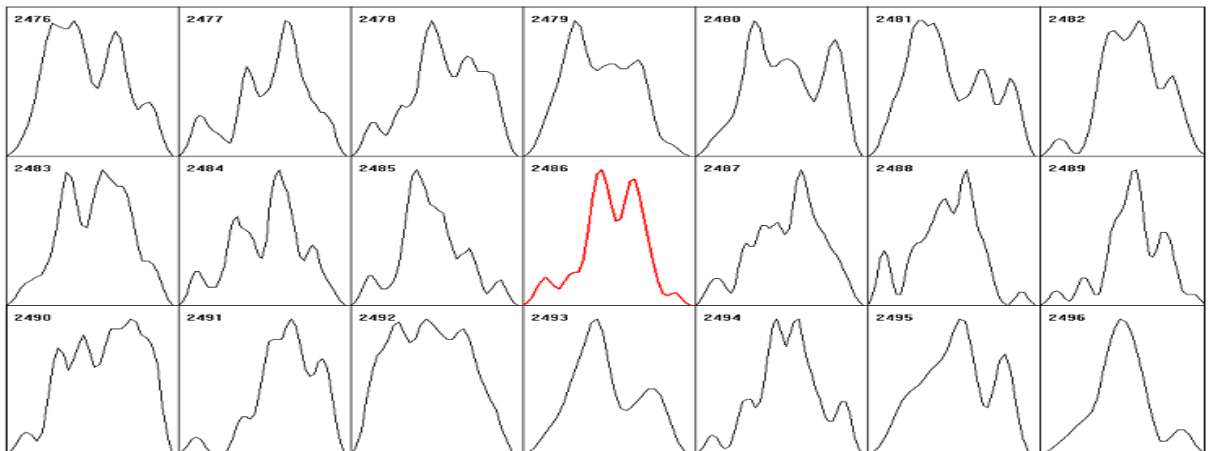

Fig. 11. 0.5-min histograms constructed from the results of measurements of noise in GCP-generator No. 103 (San Antonio TX, USA; in latitude 29°49′ north and longitude 98°62′ west) during the solar eclipse of April 8, 2005. Histogram No. 2486 (at the center of the second row) corresponds to the eclipse culmination with an accuracy of 0.5 min.

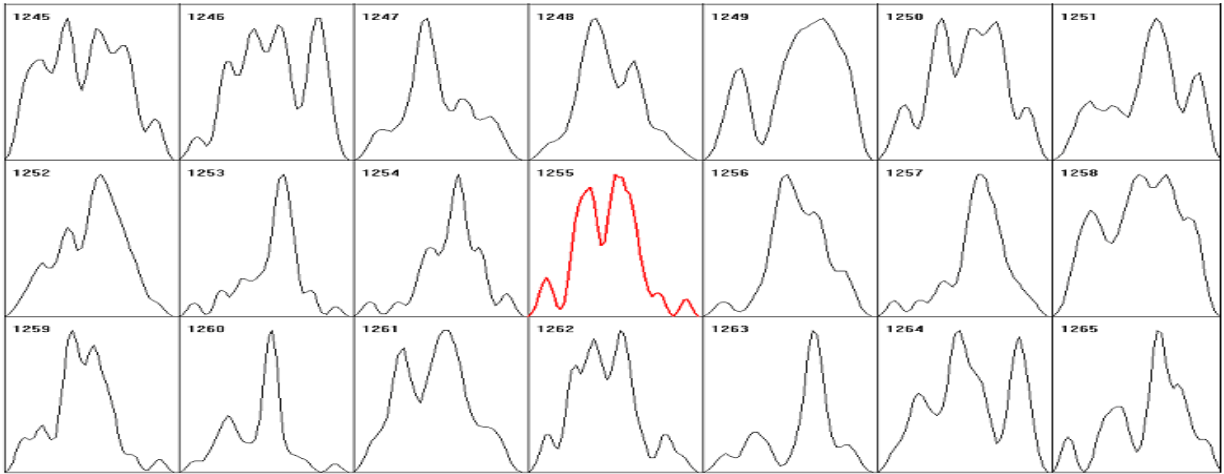

Fig. 12. 0.5-min histograms constructed from the results of measurements of noise in GCP-generator No. 28 (Princeton NJ, USA; in latitude 40°35′ north and longitude 74°66′ west) during the solar eclipse of October 3, 2005. Histogram No. 1255 (at the center of the second row) appears 0.5 min before the eclipse culmination.

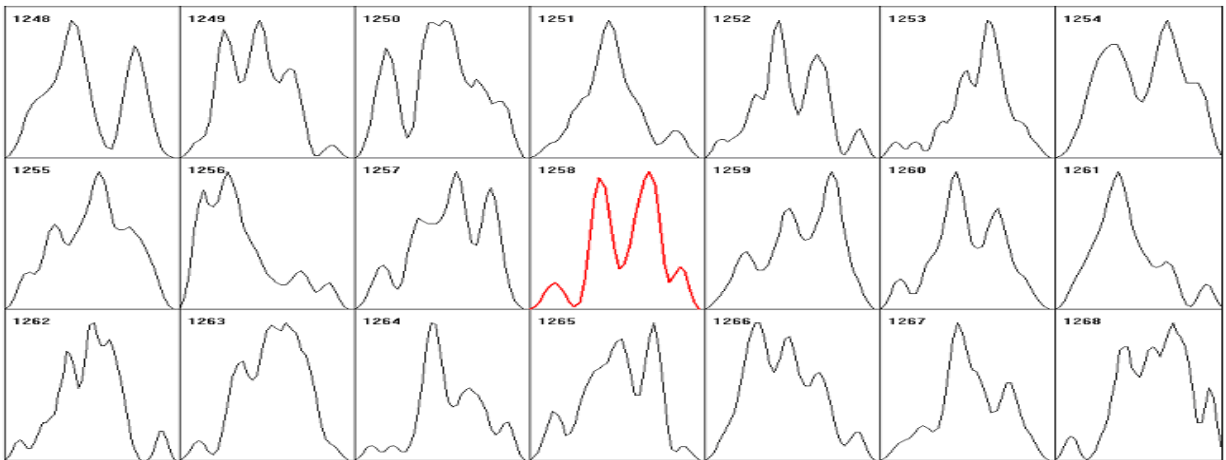

Fig. 13. 0.5-min histograms constructed from the results of measurements of noise in GCP-generator No. 37 (Neuchbtel, Switzerland; in latitude 47°08′ north and longitude 7°06′ west) during the solar eclipse of October 3, 2005. Histogram No. 1258 (at the center of the second row) appears 1 min later the eclipse culmination.

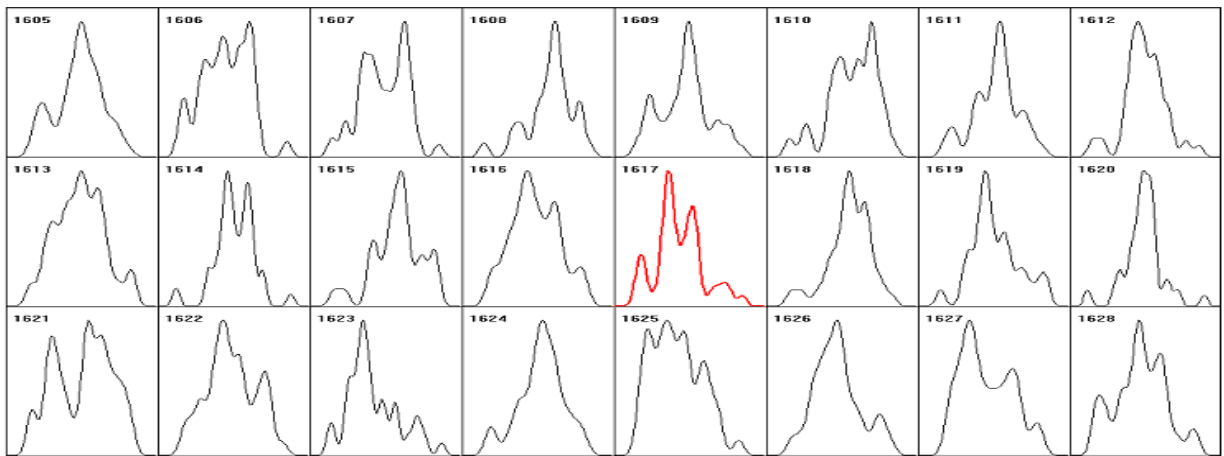

Fig. 14. 0.5-min histograms constructed from the results of measurements of $^{239}$Pu alpha-activity during the solar eclipse of October 3, 2005 in Pushchino (in latitude 54°8′ north and longitude 37°6′ east). Histogram No. 1617 (at the center of the second row) corresponds to the eclipse culmination with an accuracy of 0.5 min.

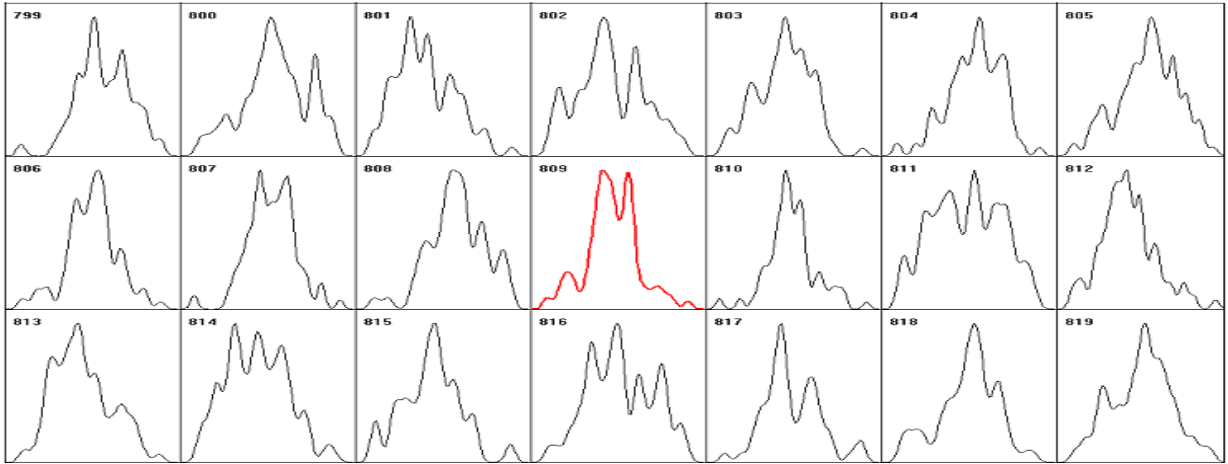

Fig. 15. 1-min histograms constructed from the results of measurements of $^{239}$Pu alpha-activity during the solar eclipse of October 3, 2005 in Pushchino (in latitude 54°8′ north and longitude 37°6′ east). Histogram No. 809 (at the center of the second row) appears 1 min later the eclipse culmination.

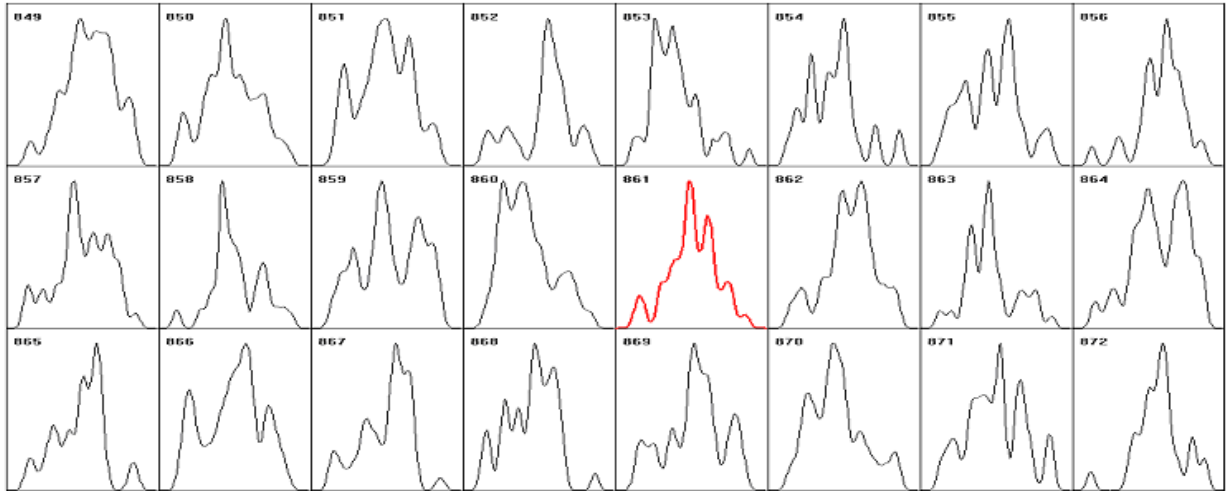

Fig. 16. 1-min histograms constructed from the results of measurements of $^{239}$Pu alpha-activity during the solar eclipse of May 31, 2003 in Antarctic (Novolazarevskaya st.; in latitude 70°02′ south and longitude 11°35′ west). Histogram No. 860 and 862, and No. 861 and 863 (at the center of the second row) correspond to the eclipse culmination in the interval of ±0.5 min.

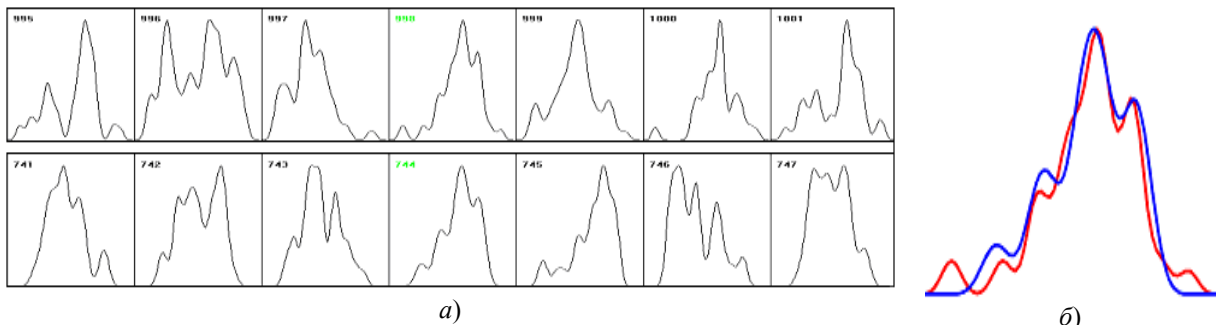

*a*)     *б*)

Fig. 17. A specific "transitional" form of 0.5-min and 1-min histograms is observed 1 min before the eclipse culminations. The upper row contains histograms constructed from the measurements of $^{239}$Pu alpha-activity in Pushchino during the solar eclipse of April 19, 2004. The second row contains analogous data obtained from the measurements during the solar eclipse of July 31, 2000. *a*) At the center of the upper and bottom rows are the histograms that appear 1 min before the eclipse culmination. *b*) The same histograms being superimposed.

Along with the histograms of the "main eclipse form", other forms, less probable but quite specific, are also observed near the culmination points of the solar eclipse. Examples of such forms are given in Fig. 17-19. Identification of such "transitional" forms is more complicated than identification of the "main eclipse form".

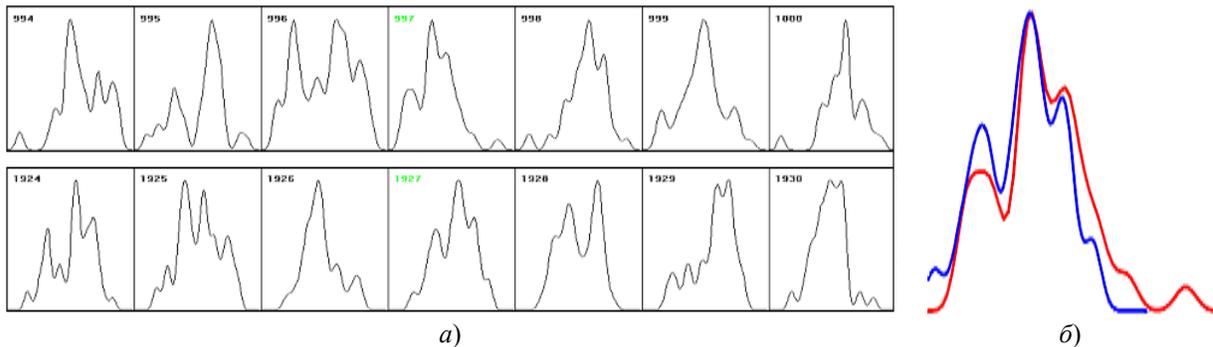

*а)* *б)*

Fig. 18. A specific "transitional" form of 0.5-min and 1-min histograms is observed 0.5-2 min before the eclipse culminations. *a)* Measurements of alpha-activity of $^{239}$Pu in Pushchino of June 21, 2001 and April 19, 2004. *b)* The histograms being superimposed.

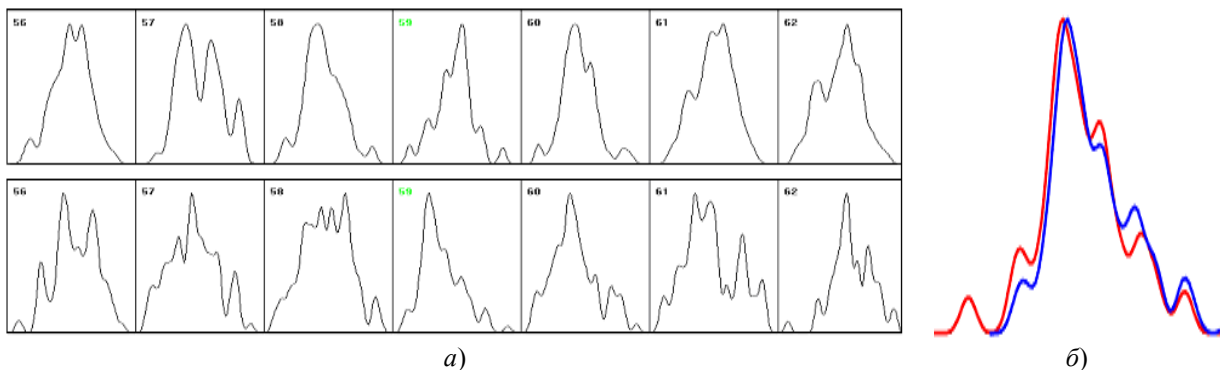

*а)* *б)*

Fig. 19. A specific "transitional" form of 0.5-min and 1-min histograms is observed 0.5-2 min before the culmination of the eclipses of December 4, 2002 and November 24, 2003. *a)* The measurements of $^{239}$Pu alpha-activity were carried out in Pushchino, with a collimator directed at the Pole Star. *b*) The histograms being superimposed.

### 4. Discussion

As can be seen from the data presented, histograms of a specific form emerge at the culmination of a solar eclipse; they are observed in different years and months, appear simultaneously (with an accuracy of 0.5-1 min) at different geographical locations and can be revealed upon measuring fluctuations in processes of various nature. The existence of a specific histogram form, corresponding to the culmination of the solar eclipse, is an example of the relation of a cosmo-geophysical process to a certain histogram form. As mentioned in the Introduction, another example of such a relation is the relation of the moment when the new moon comes to a certain histogram form, specific for this moment [9]. Let us point to characteristic features of the phenomenon described here:
- the histogram form that is specific for the moments of the new moon is clearly different from the form specific for the culmination of solar eclipses;
- the time interval, in which specific form emerge, is narrow;
- specific form appear simultaneously at different points of the globe.

It is necessary to mention that reports on "anomalous" phenomena, which accompany solar eclipses, repeatedly appeared in literature. A drastic increase of the neutron stream during the solar eclipse of July 22, 1990 was reported in [11]. The same authors also wrote about spikes in the intensity of neutron stream during the phases of the new and full moon [12]. Noteworthy are communication on changes of mass measurement of 200 gm standard laboratory weight during solar eclipse of April 8-9, 2005 made by Vezzoli [13]. Lucatelli reported about decreasing of $^{60}$Co radioactive count data rate at the moment of maximum of solar eclipse of 4 December 2002 [14]. Allais registered anomalies in the movements of a Foucault's pendulum, taking place during solar eclipses [15]. In Allen and Sax's experiments, significant variations in the period of a torsion pendulum were observed both at the moment of solar eclipse and the moments before and after it [16]. Anomalies, coinciding with the solar eclipses of September 23 1987, March 18 1988 and July 22 1990, were observed in the rhythm of an atomic clock [17]. Also, some anomalies in gravity measurements were noted at the moments of solar eclipses [18].

All works mentioned above have one point in common: registered there were changes of the mean values of measured parameters. In principle, the tide-generating forces can cause such changes. In contrast to those works, we do not observe significant changes of mean values at the culminations of solar eclipses. The subject of our analysis is the form of distribution of amplitude fluctuations of various processes, and this form is almost insensitive to the changes of mean amplitudes. More than once did we search for correlations between the phenomena discussed and various cosmo-physical factors [2, 19]. It seems that the specific histogram forms given in the present work are not related to the effect of tide-generating forces: analyzing changes of the histogram form, we have not found periods, which would correspond to these forces.

The narrowness of the interval, during which a specific form synchronously appears at different places, suggests this form to be related to a wave effect accompanying the culmination of the solar eclipse. Our previous studies indicate that histogram form would relate to fluctuations of the space-time and hence, to the gravitational interaction [8]. To sum up, we can suppose that the characteristic histogram forms described in this work result from a gravitational-wave effect, which occurs at the moment of a solar eclipse reaching its culmination.

The results of our works [20, 21] argue in favor of this supposition. As found in those works, the histogram form is sensitive to "regimes with acceleration", which can be created experimentally, by speeding up or slowing down a massive rotor. In works [22-23], which are in some way complementary to works [20, 21], used a registration system providing regimes with acceleration. It turned out, that such a system is sensitive to culminations of solar eclipses. The culmination of the solar eclipse has a relation to certain extrema in the velocity of change of the space-time position of the Sun, the Earth and the Moon; in this respect, the situation can be considered as a regime with acceleration and thus, it can determine the form of histograms describing the fluctuations in various processes.